\documentclass[11pt, a4paper]{article}

\usepackage[T2A]{fontenc}
\usepackage[utf8]{inputenc}
\usepackage[english]{babel}

\usepackage[margin=1in]{geometry} % Стандартные поля
\usepackage{microtype} % Улучшает типографику и уменьшает переносы

\usepackage{amsmath, amsfonts, amssymb, amsthm}
\usepackage{mathtools}

\usepackage[colorlinks=true, allcolors=blue]{hyperref} % Кликабельные ссылки
\usepackage{graphicx}
\usepackage{booktabs} % Красивые таблицы

\usepackage[numbers,sort&compress]{natbib} % Числовые ссылки, сжатые (напр. [1-3])

\usepackage{authblk}

\title{Linear perturbations of an exact gravitational wave\\in the Bianchi IV universe}
\author[1,2,3]{Konstantin E. Osetrin}
\affil[1]{Tomsk State Pedagogical University,\protect\\634061, Tomsk, Kievskaya str.60, Russia}
\affil[2]{Tomsk State University of Control Systems and Radioelectronics,\protect\\634050, Tomsk, Lenin str.40, Russia}
\affil[3]{National Research Tomsk State University,\protect\\634050, Tomsk, Lenin str.36, Russia}
\date{\today}

\begin{document}

\maketitle

\begin{abstract}
The proper-time method for constructing perturbative dynamical gravitational fields 
%(including gravitational wave models) 
is presented. Using the proper-time method, a perturbative analytical model of gravitational waves against the backdrop of an exact wave solution of Einstein's equations in a Bianchi~IV universe is constructed. 
To construct the perturbative analytical wave model a privileged wave coordinate system and a synchronous time function associated with the proper time of an observer freely moving in a gravitational wave were used. Reduction of the field equations, taking into account compatibility conditions, reduces the mathematical model of gravitational waves to a system of coupled ordinary differential equations for functions of the wave variable.
Analytical solutions for the components of the gravitational-wave metric have been found.
The stability of the resulting perturbative solutions is proven. The stability of the exact solution for a gravitational wave in the anisotropic Bianchi IV universe is demonstrated.
\end{abstract}

\vspace{0.5em} % Небольшой отступ
\noindent \textbf{Keywords:} exact gravitational waves, perturbative gravitational waves, proper time method, geodesic Hamilton-Jacobi equation, Bianchi universes.
\vspace{0.5em} 

\noindent \textbf{MSC 2020:} 83C35, 83C25. % Ваши коды классификации

\section{Introduction}
\label{sec1}

The emergence of gravitational-wave astronomy ten years ago in 2015 
\cite{PhysRevLett.116.061102,Abbott2017PRL161101,Abbott_2017AJL,Abbott_PhysRevX.9.011001} 
opened up a new channel, beyond electromagnetic radiation, for obtaining information about the universe, astrophysical processes, and objects. Obtaining information about nature from this new, additional channel of information dramatically expanded the possibilities of observations and theoretical research in astronomy, astrophysics, and cosmology.

Current observational data from ground-based and satellite observatories and their interpretation have posed a number of complex questions for researchers in these fields: modeling the early stages of the universe's dynamics; adequately describing the accelerated formation of galaxies, stars, and black holes; interpreting observational data of the electromagnetic microwave background (EMCB), including its anisotropy \cite{Bennett2013}; investigating the stochastic gravitational wave background (GWB) of the universe \cite{PhysRevD.75.123518,Caprini2018_163001}, and investigating the phenomena of \,''dark energy''\, and \,''dark matter'', \,among other topics. Research in these areas has reached the cutting edge of modern physical science. The role of gravitational waves in this research has greatly increased.

Gravitational waves in the early stages of the universe's dynamics \cite{MAGGIORE2000283,Saito200916,Saito2010867} could have played a significant role in the accelerated formation of inhomogeneities in the primordial plasma, dark matter, and matter, and in the formation of galaxies, stars, and black holes. Gravitational waves could have had a significant impact on the formation of the microwave electromagnetic background \cite{PhysRevLett.78.2054} and its observed anisotropy. Detection of gravitational waves and theoretical interpretation of observational data from ground-based detectors, observation of the stochastic gravitational wave background \cite{van_Remortel_2023} by both direct and indirect methods using the Pulsar Timing Array \cite{Reardon_2023,AandArefId0,Xu_2023} can provide information about many astrophysical processes. Projects for satellite gravitational wave detectors and the development of indirect methods for observing low-frequency gravitational waves are underway.

However, the theoretical basis for gravitational wave research is less well developed than that for electromagnetic radiation. The models and significance of gravitational waves in the early stages of the universe's development are poorly understood. Methods for detecting low-frequency gravitational waves, which are important for studying the early stages of the universe and the dynamics of astrophysical processes, are in their infancy.

The complexity and nonlinearity of the field equations of gravity lead to difficulties in constructing both exact and perturbative solutions for complex gravitational-wave models and in interpreting them. Nevertheless, a number of exact gravitational-wave solutions have been constructed and studied \cite{Stephani_Kramer_MacCallum_Hoenselaers_Herlt_2003,Griffiths_Podolsky_2009},  including in our papers \cite{Osetrin2022894,Osetrin1455Sym_2023,Osetrin325205JPA_2023,OSETRIN2024169619}. 
Perturbative and numerical research methods continue to develop rapidly  \cite{MUKHANOV1992203,Ma19957,BishopRezzolla2016,Domenech2021398,%
	BlanchetPostNewtonian2024,PhysRevD.109.083023,PhysRevD.110.124069}.

Detection of a gravitational wave and electromagnetic signal from the merger of two neutron stars
in 2017 \cite{Abbott2017PRL161101,Abbott_2017AJL,Abbott_PhysRevX.9.011001} 
showed with high accuracy that the speed of gravitational waves is equal to the speed of light, which has imposed restrictions on a number of theoretical models and methods used to describe gravitational waves. The use of coordinate systems with wave variables along which the spacetime interval vanishes to describe gravitational waves in theoretical models has thus received experimental confirmation.

Theoretical studies that simultaneously incorporate both electromagnetic and gravitational fields play a significant role 
\cite{Obukhov1983,Obukhov10.1142/S0219887824500920,%
	OsetrinRadiation2017,OBUKHOV2024169816,sym17091574Obukhov_2025,Osetrin2025RPJ526} 
as the basis for mutually complementary and mutually controlling sources of information on astrophysical processes from different observation channels. Work on modified gravity theories, which claim to adequately describe both the early universe and the phenomena of ''dark energy'' and ''dark matter'', is also of particular interest \cite{Odintsov2007,Odintsov2011,Capozziello2011,Odintsov2017,OsetrinSymmetry2021}. The selection of realistic theories can use gravitational wave models and observational data to evaluate them.

The discovery of anisotropy in the microwave electromagnetic background of the universe
\cite{Bennett2013}
has prompted a number of studies to interpret this fact
\cite{refId0_2023,PhysRevD.107.043513,Shi_2023}.
A number of researchers have justifiably criticized the proposed kinematic interpretation of the observed anisotropy of the electromagnetic background
\cite{refId0_2021,10.1093/mnras/stad3706}.
Given the anisotropy of the observed microwave electromagnetic background, anisotropic spacetime models \cite{OsetrinHomog2006,Obukhov10.3390/sym16101385,Obukhov202284} in the early stages of the universe's dynamics, including Bianchi universe models, are of interest, as are possible processes of their further isotropization, given the current state of the universe.
Deviations of spacetime from isotropy can significantly influence current estimates of observational data and their interpretation \cite{LUDWICK2025139717}. Gravitational waves could also have a significant influence here.

Perturbative secondary gravitational waves, as perturbations of strong gravitational waves in the early universe, can play a significant role in the accelerated formation of inhomogeneities in primordial plasma, dark matter, and matter, in the formation of the stochastic gravitational wave background of the universe, and in the isotropization processes of the universe during its dynamics.

This paper constructs one of the first analytical models of perturbative gravitational waves in the strong gravitational wave background of the Bianchi IV universe. The underlying strong gravitational wave background is an exact solution to Einstein's vacuum equations.

%%%%%%%%%%%%%%%%%%%%%%%%%%%%%%%%%%%%%%%%%%%%%%%%%%%%%%%%%%%%%%%%%%%%%%%%%%%%%%%%%%%%%%%%%%%%%

\section{The proper-time method}
\label{sec2}

When constructing both exact and perturbative models of dynamic gravitational fields (gravitational wave) against the backdrop of
exact solutions of field gravitational equations, privileged wave coordinate systems with null variables,
along which the spacetime interval vanishes, are often used.
On the other hand, the dynamic nature of the gravitational field in the models under consideration
implies the possibility of an explicit dependence of the metric on the time variable.
Therefore, in our proposed method for constructing perturbative dynamic models,
we assume that the functions of perturbative corrections to the base (background) metric depend on both the wave variable and time.
As a function of time, we propose using the proper time of free
particles on geodesics of the base (background) spacetime. Therefore, we call this method "the proper time method".
In this paper, this approach is successfully tested using the example of an exact wave solution and a perturbative  wave solution for a Bianchi~IV universe.

The construction of the time function \(\tau\) as a function of the variables of the used coordinate system can be based
on the exact or perturbative solution of the geodesic Hamilton-Jacobi equation for the test particle action function  in the background spacetime with metric~\(g^{\alpha\beta}\):
\begin{equation}
	g^{\alpha\beta}\,
	\frac{\partial\tau}{\partial x^\alpha}
	\frac{\partial\tau}{\partial x^\beta}=1
	\label{HJE}
	.\end{equation}

Constructing the complete integral of the geodesic Hamilton-Jacobi equation (\ref{HJE}) allows us to find the time function
on the geodesics of the base spacetime and makes it possible to transition (if necessary)
to a synchronous reference frame (SRF)
associated with a freely moving observer \cite{LandauEng1}. 
The observer's proper time \(\tau\) can be chosen
as the uniform time of the new synchronous reference frame, and the spatial variables can be chosen such that
the observer in the new SRF is at rest.

One of the known methods for constructing the complete integral of the geodesic Hamilton-Jacobi equation (\ref{HJE}) is the method of complete separation of variables in privileged coordinate systems, which has been developed for this equation by many authors, beginning with work by Paul St\"ackel \cite{Stackel1897145} and finally developed in the works of Vladimir Shapovalov \cite{Shapovalov1978I,Shapovalov1978II,Shapovalov1979}.

\section{Exact model of gravitational waves in the Bianchi~IV universe}
\label{sec3}

In a privileged wave coordinate system with a wave variable \(x^0\) along which the spacetime interval vanishes, the wave metric for a Bianchi space of type IV can be written in the following form \cite{OsetrinHomog2006,Osetrin2022EPJP856}:
\begin{equation}
	\begin{split}
		ds^2= &
		\,2\, dx^0dx^1
		-
		\frac{\left(x^0\right)^{2 \omega } }{\alpha  \gamma-\beta ^2}
		\Bigl[
		\left(\alpha  \log ^2({x^0})+2 \beta  \log ({x^0})+\gamma \right)
		\left({dx^2}\right)^2
		\\ &
		\mbox{}
		-
		2\, (\alpha  \log ({x^0})+\beta )
		\,
		{dx^2}{dx^3}
		+\alpha  
		\left({dx^3}\right)^2
		\Bigr] ,
		\label{MetricBianchiIV}
	\end{split}
\end{equation}
\begin{equation}
	\det g_{\alpha\beta}=-\frac{\left(x^0\right)^{4 \omega }}{\alpha  \gamma-\beta ^2 }
	,\qquad
	0<x^0
	,\qquad
	\beta ^2<\alpha  \gamma
	%	\alpha  \gamma>\beta ^2
	%	,\qquad
	%	\alpha  \gamma >0
	,\end{equation}
where the constants $\alpha$, $\beta$, $\gamma$, and $\omega$ are the parameters of the wave gravity model in Bianchi space of type IV.

The spacetime admits a covariantly constant vector $K^\alpha$, i.e., it is plane-wave spacetime:
\begin{equation}
	\nabla_\alpha K_\beta=0
	\quad
	\to
	\quad
	K_\alpha=\bigl( K_0,0,0,0 \bigr)
	,\qquad
	K_0=\mbox{const} 
\end{equation}

The homogeneity of this space is determined by the system of Killing vectors
$X_{(1)}$, $X_{(2)}$, and $X_{(3)}$, which in the wave coordinate system used can be chosen in the following form:
\begin{equation}
	X^\alpha_{(1)}=\bigl(0,0,1,0\bigr),
	\quad
	X^\alpha_{(2)}=\bigl(0,0,0,1\bigr),
	\quad
	X^\alpha_{(3)}=\bigl(-x^0,x^1,\omega x^2,\omega x^3-x^2\bigr)
	.\end{equation}

The commutation relations for the Killing vectors have the following form, defining a Bianchi space of type IV:
\begin{equation}
	\left[X_{(1)},X_{(2)}\right]=0
	,\quad
	\left[X_{(1)},X_{(3)}\right]=\omega X_{(1)}-X_{(2)}
	,\quad
	\left[X_{(2)},X_{(3)}\right]=\omega X_{(2)}
	.
\end{equation}

The wave spacetime under consideration has three independent non-zero components of the Riemann curvature tensor:
\[
\begin{split}
	{R}_{0202}  = &
	\frac{
		\left(x^0\right)^{2 \omega -2} 
	}{4 \left(\beta ^2-\alpha  \gamma \right)^2}
	\Bigl[
	-\alpha 
	\left(
	\alpha ^2-4 \alpha  \gamma  (\omega -1) \omega +4 \beta ^2 (\omega -1) \omega 
	\right)
	\log ^2({x^0}) \\
	&
	\mbox{}
	-2 
	\left(
	\alpha ^2 (\beta -4 \gamma  \omega +2 \gamma )+2 \alpha  \beta  (\beta  (2 \omega -1)-2 \gamma  (\omega -1) \omega )+4 \beta ^3 (\omega -1) \omega 
	\right)
	\log ({x^0}) \\
	&
	\mbox{}
	-4 \alpha  
	\left(
	\beta ^2+\beta  (\gamma -2 \gamma  \omega )-\gamma ^2 (\omega -1) \omega 
	\right)
	+4 \beta ^2 (-2 \beta  \omega +\beta -\gamma  (\omega -1) \omega )
	+3 \alpha ^2 \gamma
	\Bigr]\, ,
	%\end{split}
	%\]
	%	x &= 1 \,,
	%	\qquad
	%	y = 2 \,,
	%	\\
	%	z &= 3 \,.
	%\end{split}
	%$$
	%\\
	%{R}_{0202} =&
	%\frac{
		%	\left(x^0\right)^{2 \omega -2} 
		%}{4 \left(\beta ^2-\alpha  \gamma \right)^2}
	%\Bigl[
	%-\alpha 
	%\left(
	%\alpha ^2-4 \alpha  \gamma  (\omega -1) \omega +4 \beta ^2 (\omega -1) \omega 
	%\right)
	%\log ^2({x^0}) 
	%\\ &
	%\mbox{}
	%-2 
	%\left(
	%\alpha ^2 (\beta -4 \gamma  \omega +2 \gamma )+2 \alpha  \beta  (\beta  (2 \omega -1)-2 \gamma  (\omega -1) \omega )+4 \beta ^3 (\omega -1) \omega 
	%\right)
	%\log ({x^0}) 
	%\\ &
	%\mbox{}
	%-4 \alpha  
	%\left(
	%\beta ^2+\beta  (\gamma -2 \gamma  \omega )-\gamma ^2 (\omega -1) \omega 
	%\right)
	%+4 \beta ^2 (-2 \beta  \omega +\beta -\gamma  (\omega -1) \omega )
	%+3 \alpha ^2 \gamma
	%\Bigr]\, ,
	\\ 
	{R}_{0302} = &
	\frac{
		\left(x^0\right)^{2 \omega -2} 
	}{4 \left(\beta ^2-\alpha  \gamma \right)^2}
	\Bigl[
	\alpha 
	\left(
	\alpha ^2-4 \alpha  \gamma  (\omega -1) \omega +4 \beta ^2 (\omega -1) \omega 
	\right)
	\log ({x^0}) 
	\\ &
	\mbox{}
	+\alpha ^2 (\beta -4 \gamma  \omega +2 \gamma )+2 \alpha  \beta  (\beta  (2 \omega -1)-2 \gamma  (\omega -1) \omega )+4 \beta ^3 (\omega -1) \omega 
	\Bigr]\, ,
	\\ 
	{R}_{0303} = & -\frac{
		\alpha  
		\Bigl(
		\alpha ^2-4 \alpha  \gamma  (\omega -1) \omega +4 \beta ^2 (\omega -1) \omega 
		\Bigr)\,
		\left(x^0\right)^{2 \omega -2} 
	}{4 \left(\beta ^2-\alpha  \gamma \right)^2}\, .
\end{split}
\]

Einstein's vacuum equations for the metric 
%under consideration 
(\ref{MetricBianchiIV}) provide an additional condition relating the model parameters:
\begin{equation}
	\alpha ^2
	-4 \omega (\omega -1) (\beta ^2+\alpha \gamma)=0
	%+4 \alpha \gamma (\omega -1) \omega 
	%-4 \beta ^2 (\omega -1) \omega=0
	.\end{equation}
The Weyl conformal curvature tensor for the model under consideration, given admissible parameter values, cannot vanish, and the model 
%under consideration 
cannot degenerate into a conformally flat spacetime.

In the paper \cite{Osetrin2022EPJP856} for this exact wave model, solutions were found for the equations of motion of test particles and, thanks to this, a law of transformation from a privileged wave coordinate system to a synchronous reference system was found, where an observer freely moving in a gravitational wave is at rest, and the time according to the clock of this observer is used as the unified time of the new synchronous reference system \cite{LandauEng1}.

\section{Linear perturbations of an exact gravitational wave}
\label{sec4}

The background strong gravitational wave metric in a Bianchi IV universe, which is an exact solution to Einstein's vacuum equations, can be represented in a privileged wave coordinate system as follows \cite{Osetrin2022EPJP856,Osetrin2024Symmetry1456}:
\begin{equation}
	\begin{split}
		{ds}^2= & 
		\, 2\,dx^0dx^1
		+
		\left(x^0\right)^{2 \omega } 
		\left(
		\frac{
			1
		}{4\omega (\omega-1)  }
		-
		\Bigl(
		\mu+\log(x^0)
		\Bigr)^2
		\,
		\right)
		%\left(x^0\right)^{2 \omega } 
		\,\left(dx^2\right)^2
		\\
		%\]
		%\begin{equation}
		&
		\mbox{}
		+
		2\left(x^0\right)^{2 \omega} \Bigl(\mu +\log(x^0)\Bigr)
		\,
		dx^2dx^3
		-
		%\frac{\left(x^0\right)^{4 \omega }}{4\omega (1-\omega)  }
		\left(x^0\right)^{2 \omega }
		\,\left(dx^3\right)^2 ,
		\label{BaseMetricCommon}
	\end{split}
\end{equation}
\begin{equation}
	0<x^0
	,\qquad
	0<\omega<1
	.\end{equation}
Where \(\mu\) and \(\omega\) are independent parameters of the background gravitational wave, and 
\(x^0\) is the wave variable along which the spacetime interval vanishes.

%For background strong gravitational wave parameter of  
%%\(\omega= 1/4\) and 
%\(\omega= 1/2\), the field equations exhibit singularities during integration, so we will consider these cases separately.
% In this section, we will consider the most general case of secondary gravitational waves in a Bianchi~IV universe 
%for 
%%\(\omega\ne 1/4\) and 
%\(\omega\ne 1/2\).

We will seek the perturbative (secondary) gravitational wave metric in a wave coordinate system 
with the wave variable \(x^0\), which determines the background field metric of the strong gravitational wave. The spacetime interval along the wave variable vanishes. In the wave coordinate system, we have
\(
g_{00}=g_{11}=0
\).

In accordance with the perturbative approach and using the proper-time method, we seek the perturbative gravitational wave metric in the form:
\begin{equation}
g^{\alpha\beta}=g^{\alpha\beta}_{B}+\epsilon\, \Omega^{\alpha\beta}
\bigl(x^0,\tau\bigr)
\label{BaseSecondaryMetricCommon}
	,\end{equation}
where \(g^{\alpha\beta}_{B}\) is the background field metric (\ref{BaseMetricCommon}), \(\epsilon\) is the dimensionless smallness parameter (\(\epsilon\ll 1\)), \(\tau\) is the synchronous time according to the clock of an observer freely moving against the background of the basic exact wave solution 
%(\ref{BaseMetricCommon}) 
of the gravitational field equations.  
%(background exact gravitational wave found in Ref.\cite{Osetrin2022EPJP856}).

The background exact gravitational wave model was constructed in \cite{Osetrin2022EPJP856}.
%\cite{LandauEng1}
The synchronous time function \(\tau=\tau(x^0,x^1,x^2,x^3)\) for the case under consideration can be represented in the used wave coordinate system in the following form:
\begin{equation}
	\begin{split}
		\tau^2 = &
		\,
		\frac{(2 \omega -1) \left(x^2\right)^2 \left(x^0\right)^{2 \omega } 
		}{4\omega (\omega -1)  }
		\,
		\biggl[
		\,
		4 (1-2 \omega )^2 (\omega -1) \omega  \log ^2({x^0})
		\\ &
		\mbox{}
		-
		4 (\mu -1)^2 \omega 
		-1
		+
		8 \omega \left(2 \omega ^2-3 \omega +1\right)  (\mu  (2 \omega -1)+1) \log(x^0)
		\\ &
		\mbox{}
		+
		16 \mu ^2 \omega ^4
		+16 (1-2 \mu ) \mu  \omega ^3
		+4 \left(5 \mu ^2-6 \mu +1\right) \omega ^2
		\,
		\biggr]
		\\ &
		\mbox{}
		-
		2\, (2 \omega-1 )^2\, {x^2} {x^3} \left(x^0\right)^{2 \omega } \Bigl(\mu  (2 \omega -1)+1+(2 \omega -1) \log(x^0)\Bigr)
		\\ &
		\mbox{}
		+
		(2 \omega -1)^3 \left(x^3\right)^2 \left(x^0\right)^{2 \omega }
		+2\, {x^0} {x^1} \, .
	\end{split}
	\label{CommonTau}
\end{equation}

Linearization of Einstein's vacuum equations for the metric (\ref{BaseSecondaryMetricCommon}) yields very cumbersome equations, which we will not write out. A study of the compatibility of these field equations with respect to the variables \(x^2\) and \(x^3\), on which the synchronous time function \(\tau\) depends, after rather cumbersome but obvious calculations, leads to the following structure of the dependence of the components of the perturbative (secondary) gravitational wave metric on the time \(\tau\) and the wave variable~\(x^0\):
\begin{equation}
	\Omega^{00}=\Omega^{11}=0
	,\end{equation}
\begin{equation}
	\Omega^{01} = A_{01}({x^0})
	,\qquad
	\Omega^{02} = A_{02}({x^0})
	,\qquad
	\Omega^{03} = A_{03}({x^0})
	,\end{equation}
\begin{equation}
	\Omega^{12} = B_{12}({x^0})\, {\tau}^2+A_{12}({x^0})
	,\qquad
	\Omega^{13} = B_{13}({x^0})\, {\tau}^2+A_{13}({x^0})
	,\end{equation}
\begin{equation}
	\Omega^{22} = A_{22}({x^0})
	,\qquad
	\Omega^{23} = A_{23}({x^0})
	,\qquad
	\Omega^{33} = A_{33}({x^0})
	.\end{equation}

Since \(\Omega^{01}\) is included in \(g_{01}\) and depends only on \(x^0\), then by transforming the wave variable \(x^0\), we can convert \(g_{01}\) to unity and \(A_{01}({x^0})\) to zero.
The functions \(A_{12}({x^0})\) and \(A_{13}({x^0})\) are included in the components of the metric 
\(g_{02}\) and \(g_{03}\) only additively and only as functions of \(x^0\) and therefore can be set equal to zero by coordinate transformations.

The compatibility conditions for the linearized field equations contain an autonomous subsystem of two coupled first-order differential equations for the functions \(B_{12}({x^0})\) and \(B_{13}({x^0})\), which can be reduced to the following form:
\begin{equation}
	\begin{split}
		B_{12}'({x^0}) = &
		\frac{
			4 \omega  
		}{{x^0}}
		\biggl(
		B_{12}({x^0}) 
		\Bigl(
		\left(4 \omega ^3-8 \omega ^2+6 \omega -2\right) \log{(x^0)}+
		\\ &
		\mu  \left(4 \omega ^3-8 \omega ^2+6 \omega -2\right)+\omega  (2 \omega -3)
		\Bigr)
		\\ &
		\mbox{}
		+
		2 \left(1-2 \omega ^3+4 \omega ^2-3 \omega \right) B_{13}({x^0})
		\biggr)\, ,
	\end{split}
	\label{B13FromD1B12}
\end{equation}
\begin{equation}
	\begin{split}
		B_{13}'({x^0}) = &
		\frac{
			4 \omega  
		}{{x^0}}
		\biggl(
		B_{12}({x^0}) 
		\Bigl(
		\left(4 \omega ^3-8 \omega ^2+6 \omega -2\right) \log{(x^0)}
		\\ &
		+
		\mu  \left(4 \omega ^3-8 \omega ^2+6 \omega -2\right)
		+
		\omega  (2 \omega -3)
		\Bigr)
		\\ &
		\mbox{}
		+2 \left(1-2 \omega ^3+4 \omega ^2-3 \omega \right) B_{13}({x^0})
		\biggr)\, .
	\end{split}
	\label{D1B13}
\end{equation}
From here on, the superscript denotes the derivative with respect to the variable on which the function depends.

\begin{figure}
\begin{center}
\includegraphics[width=0.6\textwidth]{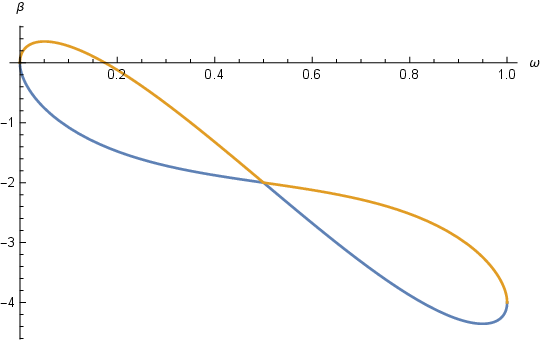}
\caption[]{
%Parameters 
\(\beta_1(\omega)\) (dark blue line) and \(\beta_2(\omega)\) (light yellow line).}
\label{GraphicBeta}
\end{center}
\end{figure}

From the system of equations (\ref{B13FromD1B12})--(\ref{D1B13}), independent equations can be obtained by increasing the order of the equations.
Thus, from equation (\ref{B13FromD1B12}), we can express the function \(B_{13}(x^0)\) in terms of
the function \(B_{12}\) (and its derivative):
\begin{equation}
	B_{13} = 
	\frac{
		4 \omega  B_{12} 
		\,
		\Bigl(
		2 (\omega -1) \bigl(2\omega (\omega -1)  +1\bigr)
		\left(\mu+ \log (x^0)\right)
		+
		\omega  (2 \omega -3)
		\Bigr)
		-{x^0} B_{12}'
	}{8 \omega  \left(2 \omega ^3-4 \omega ^2+3 \omega -1\right)}
	\label{B13EqA}
	.\end{equation}
Then, from equation (\ref{D1B13}) using relation (\ref{B13EqA}), we obtain a second-order differential equation for the function \(B_{12}(x^0)\):
\begin{equation}
	B_{12}'' = -\frac{
		(8 \omega +1) {x^0} B_{12}'
		+8 \omega \left(4 \omega ^3 - 8 \omega ^2 + 7 \omega -1\right) B_{12}
	}{\left(x^0\right)^2}
	\label{D2B12}
	.\end{equation}
The solution to equation (\ref{D2B12}) is as follows:
\begin{equation}
	B_{12}= b_1 \left(x^0\right)^{\beta_1} + b_2 \left(x^0\right)^{\beta_2}
	\label{B12Solution}
	.\end{equation}
Here \(b_1\) and \(b_2\) are constants of integration, and the parameters \({\beta_1}\) and \({\beta_2}\) are solutions of a quadratic equation of the following form:
\begin{equation}
	\beta^2+8 \omega \beta
	+8 \omega \left(4 \omega^3-8 \omega ^2+7 \omega -1\right)=0
	.\end{equation}
The parameters \(\beta_1\) and \(\beta_2\) are determined through the parameter \(\omega\) of the gravitational wave:
%\begin{equation}
%\beta_1 = 2 \left(-2 \omega-\sqrt{2} \sqrt{-4 \omega ^4+8 \omega ^3
	%-5 \omega ^2+\omega} \right)
%%\label{CommonBeta1}
%,\end{equation}
%\begin{equation} 
%\beta_2 = 2 \left(-2 \omega+\sqrt{2} \sqrt{-4 \omega ^4
%+8 \omega^3-5 \omega ^2+\omega } \right)
%%\label{CommonBeta2}
%.\end{equation}
%\begin{equation}
%\beta_1 = -4 \left(\omega+\sqrt{2}\,\sqrt{\omega (1-\omega)(\omega-1/2)^2} \right)
%\label{CommonBeta1}
%,\end{equation}
%\begin{equation}
%\beta_2 = -4 \left(\omega-\sqrt{2}\,\sqrt{\omega (1-\omega)(\omega-1/2)^2} \right)
%\label{CommonBeta2}
%.\end{equation}
\begin{equation}
\beta_1 = -4 \left(\omega+\sqrt{2\omega (1-\omega)(\omega-1/2)^2} \right)
\label{CommonBeta1}
,\end{equation}
\begin{equation}
\beta_2 = -4 \left(\omega-\sqrt{2\omega (1-\omega)(\omega-1/2)^2} \right)
\label{CommonBeta2}
.\end{equation}
The values of parameters \(\beta_1\) and \(\beta_2\) for all allowed values of the gravitational wave parameter  \(\omega\) are shown in  figure \ref{GraphicBeta}.
%The special case of \(\omega=1/2\) will be considered separately below.

Using the solution for \(B_{12}(x^0)\), we obtain the function \(B_{13}(x^0)\) 
from equation~(\ref{B13EqA}):
%\[
%B_{13}(x^0) = 
%\frac{
%\omega  (2 \omega -3)
%2
%\left(2 \omega ^3-4 \omega ^2+3 \omega -1\right) \bigl(\mu+\log(x^0)\bigr)
%}{2
%\left(2 \omega ^3-4 \omega ^2+3 \omega -1\right)}
%\times
%\]
%\begin{equation}
%\times
%\left(
%{b_1} \left(x^0\right)^{{\beta_1}}+{b_2} 
%\left(x^0\right)^{{\beta_2}}
%\right)
%-
%\frac{
%{b_1} {\beta_1} 
%\left(x^0\right)^{{\beta_1}}
%+{b_2} {\beta_2} 
%\left(x^0\right)^{{\beta_2}}
%}{8 \omega  \left(2 \omega ^3-4 \omega ^2+3 \omega -1\right)}
%.\end{equation}
%
%\[
%\begin{equation}
%B_{13}(x^0) = 
%%-
%\frac{
%{b_1} {\beta_1} 
%\left(x^0\right)^{{\beta_1}}
%+{b_2} {\beta_2} 
%\left(x^0\right)^{{\beta_2}}
%}{8 \omega  
%%\left(2 \omega ^3-4 \omega ^2+3 \omega -1\right)
%(1-\omega) \left(2 \omega ^2-2 \omega +1\right)
%}
%%
%%\]
%%\begin{equation}
%%\mbox{}
%+
%%\frac{
%\omega  (2 \omega -3)
%\left(\mu+\log(x^0)\right)
%%2
%%\left(2 \omega ^3-4 \omega ^2+3 \omega -1\right) 
%%\bigl(\mu+\log(x^0)\bigr)
%%}{2
%%\left(2 \omega ^3-4 \omega ^2+3 \omega -1\right)}
%%\times
%%\times
%\left(
%{b_1} \left(x^0\right)^{{\beta_1}}+{b_2} 
%\left(x^0\right)^{{\beta_2}}
%\right)
%\label{B13Solution}
%.\end{equation}
\begin{equation}
B_{13}(x^0) = 
%-
\frac{
{b_1} {\beta_1} 
\left(x^0\right)^{{\beta_1}}
+{b_2} {\beta_2} 
\left(x^0\right)^{{\beta_2}}
}{8 \omega  
%\left(2 \omega ^3-4 \omega ^2+3 \omega -1\right)
(1-\omega) \left(2 \omega ^2-2 \omega +1\right)
}
- \omega  (3-2 \omega)
\left(\mu+\log(x^0)\right)
B_{12}(x^0)
\label{B13Solution}
.\end{equation}

As shown in paper \cite{Osetrin2022EPJP856}, along geodesics the wave variable \(x^0\) is proportional to the synchronous time \(\tau\). 
Then the correction functions \(\Omega^{12}=\tau^2 B_{12}\) and \(\Omega^{13}=\tau^2 B_{13}\) will be limited in time when \(\beta_1<\beta_2<-2\) for the following region of parameter \(\omega\):
\begin{equation}
1/2<\omega<1
\label{omegaRegion}
.\end{equation}
Since in the admissible region of parameter \(\omega\) we have the relation \(\beta_1<\beta_2<-2\), then at large times \(\tau\) we have the following behavior
\begin{equation}
\lim_{\tau\to\infty}\Omega^{12}=
\lim_{\tau\to\infty}\tau^2 B_{12}
\to 0
,\qquad
\lim_{\tau\to\infty}\Omega^{13}=
\lim_{\tau\to\infty}\tau^2 B_{13} 
\to 0
.\end{equation}	
%\begin{equation}
%\Omega^{12}\stackrel{\longrightarrow}{\tau\to\infty}	b_2 \left(\tau\right)^{2+\beta_2}
%,\qquad
%\lim_{\tau\to\infty}\Omega^{12}=
%\lim_{\tau\to\infty}\tau^2 B_{12}
%\to	\mbox{const} \left(x^0\right)^{\beta_2}
%\to 0
%,\end{equation}
%\begin{equation}
%\Omega^{13}{x^0\to\infty} 
%,\qquad
%\lim_{\tau\to\infty}\Omega^{13}=
%\lim_{\tau\to\infty}\tau^2 B_{13} 
%\to
%b_2 \,
%\omega  (2 \omega -3)\log(x^0)
%\left(x^0\right)^{\beta_2}
%\to\omega  (2 \omega -3)\log(x^0) B_{12}
%\to 0
%
%\left(
%\frac{
%{\beta_2}
%}{8 \omega  
%	%\left(2 \omega ^3-4 \omega ^2+3 \omega -1\right)
%	(1-\omega) \left(2 \omega ^2-2 \omega +1\right)
%}
%+
%\omega  (2 \omega -3)
%\Bigl(\mu+\log(x^0)\Bigr)
%\left(
%{b_1} \left(x^0\right)^{{\beta_1}}+{b_2} 
%\left(x^0\right)^{{\beta_2}}
%\right)
%\right)
%.\end{equation}
From the compatibility conditions of the field equations, we additionally obtain a coupled system of two second-order differential equations. For the function \(A_{02}(x^0)\) we have 
%and \(A_{03}(x^0)\) of the following form:
\begin{equation}
\begin{split}
A_{02}'' = &
\frac{
2 
}{{x^0}}
\biggl(
\,
A_{02}'
\,
\Bigl(
2 \mu  \omega ^2 -2 \mu  \omega  -2 \omega  
+2 \omega\bigl(\omega-1\bigr) \log{(x^0)}
\Bigr)
\\ &
\mbox{}
+
2 \omega
\bigl(
1- \omega 
\bigr)
A_{03}'
+\left(x^0\right)^2 B_{12}'({x^0})
\\ &
\mbox{}
-
2 {x^0} B_{12}
\Bigl(
4\omega\left(2\omega^3-4\omega^2+3\omega -1\right) \log{(x^0)}
\\ &
\mbox{}
+
8 \mu  \omega ^4+(4-16 \mu ) \omega ^3+6 (2 \mu -1) \omega ^2-4 \mu  \omega 
+
\omega -1
\Bigr)
\\ &
\mbox{}
+8\omega \left(2 \omega ^3-4 \omega ^2+3 \omega -1\right) 
{x^0} B_{13}
\biggr)\, .
\end{split}
\label{ComonEqForA02byA02}
\end{equation}
and for the function \(A_{03}(x^0)\) we have the following equation:
\begin{equation}
\begin{split}
A_{03}'' = &
\frac{
1
}{{x^0}}
\Biggl(
A_{02}'
\,
\biggl(
4 \mu ^2 \omega
\left(
\omega -1  
\right)
+1
+
8 \mu  \omega 
\bigl( \omega 
- 1
\bigr)
\log{(x^0)} 
+
4 \omega
\bigl(
\omega 
-1  
\bigr)
\log^2(x^0) 
\biggr)
\\ & 
\mbox{}
+
A_{03}'
\,
\biggl(
4 \mu  \omega  
-4 \mu  \omega ^2 
-4 \omega ^2 \log{(x^0)} 
-4 \omega  
+4 \omega  \log{(x^0)} 
\biggr)
\\ & 
\mbox{}
+2 \left(x^0\right)^2 
B_{13}'
+4 {x^0} B_{13}
\Bigl(
4\omega  \left(2 \omega ^3-4 \omega ^2+3 \omega -1\right) \log{(x^0)}
\\ & 
\mbox{}
+
8 \mu  \omega ^4+(4-16 \mu ) \omega ^3+6 (2 \mu -1) \omega ^2+(3-4 \mu ) \omega 
+1
\Bigr)
\\ & 
\mbox{}
-4 {x^0} B_{12}
\Bigl(
4\omega  \left(2 \omega ^3-4 \omega ^2+3 \omega -1\right)  \log^2(x^0)
\\ & 
\mbox{}
+
4  \omega  (\omega -1)
\left(\mu \left(4 \omega ^2-4 \omega +2\right)+2 \omega -1\right) \log{(x^0)}
\\ &
\mbox{}
+
8 \mu ^2 \omega ^4
+
2 \omega ^2 \left(6 \mu ^2-6 \mu +1\right)
+
\omega \left(-4 \mu ^2+4 \mu -2\right)  
+8\omega ^3 \mu (1-2 \mu )   
+1
\Bigr)
\Biggr)\, .
\end{split}
\label{CommonEqD2A03}
\end{equation}

Expressing the derivative of \(A_{03}'\) from equation (\ref{ComonEqForA02byA02}), taking into account the form of the previously found derivatives for the functions \(B_{12}'\) and \(B_{13}'\), we obtain:
\begin{equation}
A_{03}' =
\frac{
{x^0} A_{02}''+4 \omega A_{02}' \left(
(1 -\omega )\left(\mu+ \log(x^0)\right)+1\right)
}{4\omega (1-\omega) }
+
\frac{
{x^0} B_{12}({x^0})
}{ \omega }
\label{CommonD1A03byA02}
.\end{equation}
This equation determines the first derivative of function \(A_{03}\) with respect to functions \(A_{02}\) and \(B_{12}\), which are assumed to be given.

Substituting the derivative \(A_{03}'\) from eq.~(\ref{CommonD1A03byA02}) to eq.~(\ref{CommonEqD2A03}) we obtain a third-order differential equation for determining the function \(A_{02}(x^0)\):
\begin{equation}
\begin{split}
A_{02}''' = &
-\frac{(8 \omega +1) A_{02}''({x^0})}{{x^0}}-\frac{16 \omega ^2 A_{02}'({x^0})}{\left(x^0\right)^2}
\\ &
\mbox{}
+
\frac{4 (1-\omega) B_{12}({x^0}) 
}{{x^0}}
\Bigl(
4\omega \left(4 \omega ^3-8 \omega ^2+7 \omega -3\right) 
\log({x^0})
\\ &
\mbox{}
+
16 \mu  \omega ^4+(8-32 \mu ) \omega ^3+4 (7 \mu -3) \omega ^2+(4-12 \mu ) \omega 
+
1
\Bigr)
\\ &
\mbox{}
+\frac{16 \omega  \left(4 \omega ^2-4 \omega +3\right) (\omega -1)^2 B_{13}({x^0})
}{{x^0}} \, .
\end{split}
\label{CommonEqD3A02}
\end{equation}
We obtain a third-order linear inhomogeneous differential equation with variable coefficients, where the inhomogeneous part is defined by functions \( B_{12}\) and \(B_{13}\) from eq.~(\ref{B12Solution}) and eq.~(\ref{B13Solution}).

Integrating equation (\ref{CommonEqD3A02}) taking into account the explicit form of the previously found functions \(B_{12}({x^0})\) and \(B_{13}({x^0})\), we obtain a solution in the following form:
\begin{equation}
\begin{split}
A_{02}(x^0) = &
\,
\frac{{a_1} \left(x^0\right)^{1-4 \omega }}{1-4 \omega }-\frac{4 {a_2} \omega \left(x^0\right)^{1-4 \omega }}{(1-4 \omega )^2}+\frac{4 {a_2} \omega \left(x^0\right)^{1-4 \omega } \log(x^0)}{1-4 \omega }+{a_3}
\\ &
-\frac{2 {b_1} (\omega -1) \left({\beta_1} \left(4 \omega ^2-4 \omega +3\right)+8 \omega ^3+4 \omega +2\right)
}{({\beta_1}+2) \left(2 \omega ^2-2 \omega +1\right) ({\beta_1}+4 \omega +1)^2}
\, 
\left(x^0\right)^{{\beta_1}+2}
\\ &
-\frac{2 {b_2} (\omega -1) \left({\beta_2} \left(4 \omega ^2-4 \omega +3\right)+8 \omega ^3+4 \omega +2\right) }{({\beta_2}+2) \left(2 \omega ^2-2 \omega +1\right) ({\beta_2}+4 \omega +1)^2}
\,
\left(x^0\right)^{{\beta_2}+2}\, ,
\end{split}
\label{CommonEqD3A03}
\end{equation}
where \(a_1\), \(a_2\), and \(a_3\) are new constants of integration.
It is easy to see that 
for region \(1/2<\omega<1\) \,(\(\beta_1<\beta_2<-2\)), the function \(A_{02}\) is a bounded function of time.
%
%The solution we obtained for \(A_{02}(x^0)\) has a singularity at the parameter value \(\omega=1/4\).
%
%Therefore, the case of a secondary gravitational wave
%for the parameter value \(\omega=1/4\) will be considered separately below.

The solution for \(A_{02}(x^0)\) allows us to obtain the function \(A_{03}(x^0)\) from equation (\ref{CommonD1A03byA02}):
\begin{align}
A_{03}(x^0) = &
\mbox{}\,
{a_4}
-
\frac{
1
}{2 (\omega -1) \omega }
\,
\Biggl[
\frac{
8 {a_2} \omega ^2 \left(x^0\right)^{1-4 \omega }
}{(1-4 \omega )^2}
+
\frac{
8 {a_2} \omega ^2 \left(x^0\right)^{1-4 \omega } \log(x^0)
}{4 \omega -1}
\nonumber \\ 
&
\mbox{}
+
\frac{2 \omega  \left(x^0\right)^{1-4 \omega } 
}{(4 \omega -1)^3}
\Biggl(
4 {a_2}\omega (1-4 \omega )^2 (\omega -1) 
\log ^2({x^0})
%{a_1} (4 \omega -1) 
%\Bigl(
%4 \mu  \omega ^2-5 \mu  \omega +\mu -3 \omega 
%\Bigr)
%+4 {a_2} \omega  
%\Bigl(
%4 \mu  \omega ^2-5 \mu  \omega +\mu -2 \omega -1
%\Bigr)
%\end{align}
%\end{equation}
%\begin{equation}
%\begin{align}
\nonumber \\ 
&
\mbox{}
+
(4 \omega -1)
\Bigl(
{a_1} 
\left(4 \omega ^2-5 \omega +1\right)
+4 {a_2} \omega  
\left(4 \mu  \omega ^2-5 \mu  \omega +\mu -2 \omega -1\right)
\Bigr)
\, \log(x^0) 
\nonumber \\ 
&
\mbox{}
+
{a_1} (4 \omega -1) 
\Bigl(
4 \mu  \omega ^2-5 \mu  \omega +\mu -3 \omega 
\Bigr)
+4 {a_2} \omega  
\Bigl(
4 \mu  \omega ^2-5 \mu  \omega +\mu -2 \omega -1
\Bigr)
\Biggr)
%\end{align}
%\]
\nonumber \\ 
% \displaybreak
&
%\[
%\end{align}
%\begin{align}
\mbox{}
+
\frac{
2 {a_1} \omega  \left(x^0\right)^{1-4 \omega }
}{4 \omega -1}
+
\frac{
2 {a_2} \omega  \left(x^0\right)^{1-4 \omega }
}{1-4 \omega }
%\nonumber \\ 
%&
%\mbox{}
+
\frac{
2 {b_1} (\omega -1) \left(x^0\right)^{{\beta_1}+2}
}{{\beta_1}+2}
+
\frac{
2 {b_2} (\omega -1) \left(x^0\right)^{{\beta_2}+2}
}{{\beta_2}+2}
\nonumber \\ 
&
\mbox{}
+
\frac{
4 {b_1} (\omega -1) \omega  \left({\beta_1} \left(4 \omega ^2-4 \omega +3\right)+8 \omega ^3+4 \omega +2\right) 
}{({\beta_1}+2)^2 \left(2 \omega ^2-2 \omega +1\right) ({\beta_1}+4 \omega +1)^2}
\times
\nonumber \\
&
\times
\Bigl({\beta_1} (\mu  (\omega -1)-1)
+2 \mu  \omega -2 \mu -\omega -1
+({\beta_1}+2) (\omega -1) \log(x^0)
\Bigr)
\left(x^0\right)^{{\beta_1}+2} 
\nonumber \\ 
&
\mbox{}
+
\frac{
4 {b_2} (\omega -1) \omega  
\left(
{\beta_2} \left(4 \omega ^2-4 \omega +3\right)+8 \omega ^3+4 \omega +2
\right) 
}{({\beta_2}+2)^2 \left(2 \omega ^2-2 \omega +1\right) ({\beta_2}+4 \omega +1)^2}
\times
\nonumber \\ &
\times
\Bigl({\beta_2} (\mu  (\omega -1)-1)
+2 \mu  \omega -2 \mu -\omega -1
+({\beta_2}+2) (\omega -1) \log(x^0)
\Bigr)
\left(x^0\right)^{{\beta_2}+2} 
\nonumber \\ &
\mbox{}
-
\frac{
{b_1} ({\beta_1}+1) (\omega -1) \left({\eta_1} \left(4 \omega ^2-4 \omega +3\right)+8 \omega ^3+4 \omega +2\right)
}{({\beta_1}+2) \left(2 \omega ^2-2 \omega +1\right) ({\beta_1}+4 \omega +1)^2}
\,
\left(x^0\right)^{{\beta_1}+2}
\nonumber \\ &
\mbox{}
-
\frac{
{b_2} ({\beta_2}+1) (\omega -1) \left({\beta_2} \left(4 \omega ^2-4 \omega +3\right)+8 \omega ^3+4 \omega +2\right) 
}{({\beta_2}+2) \left(2 \omega ^2-2 \omega +1\right) ({\beta_2}+4 \omega +1)^2}
\,
\left(x^0\right)^{{\beta_2}+2}
\,
\Biggr]\, .
\end{align}
%\]
It is easy to see that for the region of \(1/2<\omega<1\), where \(\beta_1<\beta_2<-2\), the function \(A_{03}\) is also time-bounded.

From the system of field equations, only one equation remains, connecting the three remaining undefined functions \(A_{22}(x^0)\), \(A_{23}(x^0)\) and \(A_{33}(x^0)\):
\[
%\begin{equation}
\begin{split}
\left(x^0\right)^{2 \omega +2} A_{22}'' 
\,
\biggl(
&
4 \mu ^2 (1-\omega) \omega + 1 +8 \mu  (1-\omega ) \omega  \log(x^0)+4 (1-\omega ) \omega  \log ^2({x^0})
\biggr)
\\ &
\mbox{}
+
6 \omega  \left(x^0\right)^{2 \omega +1} A_{22}' 
\biggl(
\mu ^2 (1-\omega ) \omega+4 \mu  (1-\omega )+1 
\\ &
\mbox{}
+
4 (1-\omega) (2 \mu  \omega +1) \log(x^0)
+
4 (1-\omega ) \omega  \log ^2({x^0})
\biggr)
\end{split}
\]
\[
%\begin{equation}
%\begin{split}
\mbox{}
-
2 \omega 
\left(x^0\right)^{2 \omega } 
A_{22}
\,
\biggl(
4 \mu ^2 \omega  \left(2 \omega ^2-\omega -1\right)
+
4 \mu  \left(8 \omega ^2-9 \omega +1\right)
+2 \omega -5
%\\ &
\]
\[
\mbox{}
+
4 (\omega -1) \left(4 \mu  \omega ^2+2 (\mu +4) \omega -1\right) \log(x^0)
+
4 \omega  \left(2 \omega ^2-\omega -1\right) \log ^2({x^0})
\biggr)
\]
\[
\mbox{}
+
8 (\omega -1) \omega  \left(x^0\right)^{2 \omega +2} A_{23}'' 
\Bigl(\mu +\log(x^0)\Bigr)
\]
\[
\mbox{}
+
24 (\omega -1) \omega  \left(x^0\right)^{2 \omega +1} A_{23}' 
\Bigl(2 \mu  \omega+1 +2 \omega  \log(x^0)\Bigr)
\]
\[
\mbox{}
+
8 (\omega -1) \omega  A_{23} 
\biggl(
%&
4 \mu  \omega ^2+2 (\mu +4) \omega-1 +2 (2 \omega +1) \omega  \log(x^0)
\biggr)
\left(x^0\right)^{2 \omega } 
\]
\[
\mbox{}
+
4 (1-\omega ) \omega  \left(x^0\right)^{2 \omega +2} A_{33}''
+
24 (1-\omega) \omega ^2 \left(x^0\right)^{2 \omega +1} A_{33}'
\]
\begin{equation}
\mbox{}
-
8 \omega ^2 \left(2 \omega ^2-\omega -1\right) 
\left(x^0\right)^{2 \omega }
A_{33} 
=0\, .
%\end{split}
\label{CommonEqForA22A23A33}
\end{equation}
Thus, the solution for the gravitational wave metric contains two arbitrary functions 
from the three functions: 
\(A_{22}(x^0)\), \(A_{23}(x^0)\), and \(A_{33}(x^0)\).
Note that for any parameter \(\omega\), 
we have a trivial particular 
solution \(A_{22}=A_{23}=A_{33}=0\).
%For any \(\omega\), equation (\ref{CommonEqForA22A23A33}) has a trivial solution of the form \(A_{22}=A_{23}=A_{33}=0\).

Let us give an example of a nonzero time-limited solution for eq.~(\ref{CommonEqForA22A23A33}) of the following form:
\begin{equation}
A_{22}=A_{23}=0
,\end{equation}
\begin{equation}
A_{33}(x^0)=c_1 (x^0)^{\gamma_1} +c_2 (x^0)^{\gamma_2} 
,\qquad
\gamma_{1,2}=\frac{1}{2}\left(1-6\omega\, \pm\sqrt{1-20\,\omega (1-\omega) } \right)
\label{exampleA33}
,\end{equation}
where \(c_1\) and \(c_2\) are integration constants. 
The solution (\ref{exampleA33}) will be time-bounded for the~following region of the parameter \(\omega\):
\begin{equation}
\frac{1}{2}+\frac{1}{\sqrt{5}}<\omega<1
%,\qquad
%\frac{1}{2}+\frac{1}{\sqrt{5}}\approx 0.947214
.\end{equation}

We have thus considered all linearized field equations for all functions included in the metric of the perturbative wave model based on the exact gravitational wave solution for a Bianchi type IV universe. We have shown that time-bounded solutions exist for all functions of the metric.

Let's write out the final form of the components of the perturbative gravitational wave metric in a Bianchi type IV universe for the most general variant of the wave parameter values considered:
\[
%\begin{equation}
1/2<\omega<1
%,\qquad
%\omega\ne 1/4
%,\qquad 
%\omega\ne 1/2
%.\end{equation}
.\]
\begin{equation}
%\begin{align}
g_{00} = 
0
,\qquad
g_{01} = 1
,\qquad
g_{11} =0,
%\end{align}
\end{equation}
%\nonumber 
%\\
%,\end{equation}
%\[
%\begin{subequations}
%\begin{align}
\begin{equation}
\begin{split}
%\begin{align}
g_{02} =   &
\,
\epsilon
{\tau}^2 \,
\Biggl[ 
\,
\frac{
\bigl(\mu +\log(x^0)\bigr) 
\left(
{b_1} {\beta_1} \left(x^0\right)^{\beta_1+2 \omega}
+{b_2} {\beta_2} \left(x^0\right)^{\beta_2+2 \omega}
\right)
}{8\omega (\omega -1)   \left(2 \omega ^2-2 \omega +1\right)}
%\nonumber 
\\ &
\mbox{}
+
\frac{
\Bigl(
4 \mu  \omega ^3
+(2-6 \mu ) \omega ^2
-2 \omega +1
+
2\omega ^2 (2 \omega -3)  \log(x^0)
\Bigr)
}{4 \omega  (1-\omega ) \left(2 \omega ^2-2 \omega +1\right)}
\times
%\nonumber 
\\ &
\times
\left({b_1}\! \left(x^0\right)^{{\beta_1}+2 \omega}
+{b_2}\!  \left(x^0\right)^{{\beta_2}+2 \omega}
\right)
\,
\Biggr]\, ,
%\end{split}
%\end{align}
\end{split}
\end{equation}
%\[
\begin{equation}
\begin{split}
%\begin{align}
g_{03} = &
\,
\epsilon  
{\tau}^2 \,
\Biggl[ 
\,
\frac{\omega  (2 \omega -3) 
%\left(x^0\right)^{2 \omega } 
\left({b_1} \left(x^0\right)^{{\beta_1}+2 \omega}+{b_2} \left(x^0\right)^{{\beta_2}+2 \omega}\right)
}{4 \omega ^3-8 \omega ^2+6 \omega -2}
%\]
%\begin{equation}
%\nonumber 
\\ &
\mbox{}
-
\frac{
%\left(x^0\right)^{2 \omega } \left(
{b_1} {\beta_1} \left(x^0\right)^{{\beta_1}+2 \omega}+{b_2} {\beta_2} \left(x^0\right)^{{\beta_2}+2 \omega}
%\right)
}{8 \omega  \left(2 \omega ^3-4 \omega ^2+3 \omega -1\right)}
\,
\Biggr]\, ,
%\end{equation}
%\end{align}
\end{split}
\end{equation}
%\end{subequations}
%\[
%\begin{equation}
%\begin{split}
\begin{align}
g_{12} = &
-\frac{\epsilon  \left(x^0\right)^{2 \omega } 
}{4 (\omega -1) \omega }
\Biggl(
\Bigl(
1
-4 \mu ^2 (\omega -1) \omega  -8 \mu  (\omega -1) \omega  \log(x^0)
-4 (\omega -1) \omega  \log ^2({x^0})
\Bigr)
\times
\nonumber \\ &
\times
\biggl(
\frac{{a_1} \left(x^0\right)^{1-4 \omega }}{1-4 \omega }
-
\frac{4 {a_2} \omega  \left(x^0\right)^{1-4 \omega }}{(1-4 \omega )^2}
+
\frac{4 {a_2} \omega  \left(x^0\right)^{1-4 \omega } \log(x^0)}{1-4 \omega }
+ {a_3}
\nonumber \\ &
\mbox{}
-
\frac{2 {b_1} (\omega -1) \left({\beta_1} \left(4 \omega ^2-4 \omega +3\right)+8 \omega ^3+4 \omega +2\right) 
}{({\beta_1}+2) \left(2 \omega ^2-2 \omega +1\right) ({\beta_1}+4 \omega +1)^2}
\,
\left(x^0\right)^{{\beta_1}+2}
\nonumber \\ &
\mbox{}
-
\frac{2 {b_2} (\omega -1) \left({\beta_2} \left(4 \omega ^2-4 \omega +3\right)+8 \omega ^3+4 \omega +2\right) 
}{({\beta_2}+2) \left(2 \omega ^2-2 \omega +1\right) ({\beta_2}+4 \omega +1)^2}
\,
\left(x^0\right)^{{\beta_2}+2}
\biggr)
\nonumber \\ 
&
\mbox{}
+
4 (\omega -1) \omega  \Bigl(\mu +\log(x^0)\Bigr) 
\biggl[
{a_4}
+
\frac{{x^0} 
}{2\omega (1-\omega)  }
\biggl(
\frac{2 {a_1} \omega  \left(x^0\right)^{-4 \omega }
}{4 \omega -1}
+
\frac{8 {a_2} \omega ^2 \left(x^0\right)^{-4 \omega }
}{(1-4 \omega )^2}
\nonumber \\ 
% \displaybreak
&
\mbox{}
+
\frac{
2 \omega  \left(x^0\right)^{-4 \omega } 
}{(4 \omega -1)^3}
\Bigl(
{a_1} (4 \omega -1) 
\left(
4 \mu  \omega ^2-5 \mu  \omega +\mu -3 \omega 
\right)
\nonumber \\ 
&
\mbox{}
+
4 {a_2} \omega  \left(4 \mu  \omega ^2-5 \mu  \omega +\mu -2 \omega -1\right)
\nonumber \\ &
\mbox{}
+
(4 \omega -1) 
\left(
{a_1} \left(4 \omega ^2-5 \omega +1\right)
+
4 {a_2} \omega  \left(4 \mu  \omega ^2-5 \mu  \omega +\mu -2 \omega -1\right)
\right)
\log(x^0) 
\nonumber \\ &
\mbox{}
+
4 {a_2} (1-4 \omega )^2 (\omega -1) \omega  \log ^2({x^0})
\Bigr)
+
\frac{8 {a_2} \omega ^2 \left(x^0\right)^{-4 \omega } \log(x^0)
}{4 \omega -1}
+
\frac{2 {a_2} \omega  \left(x^0\right)^{-4 \omega }
}{1-4 \omega }
\nonumber \\ &
\mbox{}
+
\frac{
4 {b_1} (\omega -1) \omega  
\left({\beta_1} \left(4 \omega ^2-4 \omega +3\right)+8 \omega ^3+4 \omega +2\right) 
}{({\beta_1}+2)^2 \left(2 \omega ^2-2 \omega +1\right) ({\beta_1}+4 \omega +1)^2}
\times
\nonumber \\ &
\times
\Bigl(
{\beta_1} (\mu  (\omega -1)-1)
+2 \mu  \omega -2 \mu -\omega -1
+({\beta_1}+2) (\omega -1) \log(x^0)
\Bigr)
\,
\left(x^0\right)^{{\beta_1}+1} 
\nonumber \\ &
\mbox{}
+
\frac{4 {b_2} (\omega -1) \omega  \left({\beta_2} \left(4 \omega ^2-4 \omega +3\right)+8 \omega ^3+4 \omega +2\right)
}{({\beta_2}+2)^2 \left(2 \omega ^2-2 \omega +1\right) ({\beta_2}+4 \omega +1)^2}
\times
\nonumber \\ &
\times
\Bigl(
{\beta_2} (\mu  (\omega -1)-1)+2 \mu  \omega -2 \mu -\omega -1
+({\beta_2}+2) (\omega -1) \log(x^0)
\Bigr)
\,
\left(x^0\right)^{{\beta_2}+1} 
\nonumber \\ &
\mbox{}
+\frac{
2 {b_1} (\omega -1) \left(x^0\right)^{{\beta_1}+1}
}{{\beta_1}+2}
+
\frac{2 {b_2} (\omega -1)\left(x^0\right)^{{\beta_2}+1}
}{{\beta_2}+2}
\nonumber \\ &
\mbox{}
-
\frac{
{b_1} ({\beta_1}+1) (\omega -1) 
\left(
{\beta_1} \left(4 \omega ^2-4 \omega +3\right)+8 \omega ^3+4 \omega +2
\right) 
}{({\beta_1}+2) \left(2 \omega ^2-2 \omega +1\right) ({\beta_1}+4 \omega +1)^2}
\,
\left(x^0\right)^{{\beta_1}+1}
\nonumber \\ &
\mbox{}
-
\frac{
{b_2} ({\beta_2}+1) (\omega -1) \left({\beta_2} \left(4 \omega ^2-4 \omega +3\right)+8 \omega ^3+4 \omega +2
\right) 
}{({\beta_2}+2) \left(2 \omega ^2-2 \omega +1\right) ({\beta_2}+4 \omega +1)^2}
\,
\left(x^0\right)^{{\beta_2}+1}
\biggr)
\biggr]
\Biggr)\, ,
%\end{split}
%\end{equation}
\end{align}
%\[
%\begin{equation}
%\begin{split}
\begin{align}
g_{13} = &
\,
\epsilon  
\left(x^0\right)^{2 \omega } 
\biggl(
-(\mu +\log(x^0)) 
\biggl[
\frac{{a_1} \left(x^0\right)^{1-4 \omega }}{1-4 \omega }
-
\frac{4 {a_2} \omega  \left(x^0\right)^{1-4 \omega }
}{(1-4 \omega )^2}
\nonumber \\ &
\mbox{}
+
\frac{4 {a_2} \omega  \left(x^0\right)^{1-4 \omega } \log(x^0)
}{1-4 \omega }
+
{a_3}
\nonumber \\ &
\mbox{}
-
\frac{2 {b_1} (\omega -1) \left({\beta_1} \left(4 \omega ^2-4 \omega +3\right)+8 \omega ^3+4 \omega +2\right) 
}{({\beta_1}+2) \left(2 \omega ^2-2 \omega +1\right) ({\beta_1}+4 \omega +1)^2}
\,
\left(x^0\right)^{{\beta_1}+2}
\nonumber \\ &
\mbox{}
-
\frac{
2 {b_2} (\omega -1) \left({\beta_2} \left(4 \omega ^2-4 \omega +3\right)+8 \omega ^3+4 \omega +2\right) 
}{({\beta_2}+2) \left(2 \omega ^2-2 \omega +1\right) ({\beta_2}+4 \omega +1)^2}
\,
\left(x^0\right)^{{\beta_2}+2}
\,
\biggr]
\nonumber \\ 
&
\mbox{}
-
\frac{{x^0} 
}{2 (\omega -1) \omega }
\,
\biggl[
{a_4}
+
\frac{2 {a_1} \omega  \left(x^0\right)^{-4 \omega }
}{4 \omega -1}
+
\frac{8 {a_2} \omega ^2 \left(x^0\right)^{-4 \omega }
}{(1-4 \omega )^2}
+
\frac{8 {a_2} \omega ^2 \left(x^0\right)^{-4 \omega } \log(x^0)
}{4 \omega -1}
\nonumber \\ &
\mbox{}
+
\frac{2 {a_2} \omega  \left(x^0\right)^{-4 \omega }
}{1-4 \omega }
+
\frac{2 {b_1} (\omega -1) \left(x^0\right)^{{\beta_1}+1}
}{{\beta_1}+2}
+
\frac{2 {b_2} (\omega -1) \left(x^0\right)^{{\beta_2}+1}
}{{\beta_2}+2}
\nonumber \\ &
\mbox{}
+
\frac{2 \omega  
\left(x^0\right)^{-4 \omega } 
}{(4 \omega -1)^3}
\biggl(
{a_1} (4 \omega -1) 
\left(
4 \mu  \omega ^2-5 \mu  \omega +\mu -3 \omega 
\right)
\nonumber \\ &
\mbox{}
+
(4 \omega -1)
\Bigl(
{a_1} \left(4 \omega ^2-5 \omega +1\right)+4 {a_2} \omega  \left(4 \mu  \omega ^2-5 \mu  \omega +\mu -2 \omega -1\right)
\Bigr)
\log(x^0) 
\nonumber \\ &
\mbox{}
+4 {a_2} \omega  
\left(
4 \mu  \omega ^2-5 \mu  \omega +\mu -2 \omega -1
\right)
+
4 {a_2} (1-4 \omega )^2 (\omega -1) \omega  \log ^2({x^0})
\biggr)
\nonumber \\ &
\mbox{}
+
\frac{4 {b_1} (\omega -1) \omega  \left({\beta_1} \left(4 \omega ^2-4 \omega +3\right)+8 \omega ^3+4 \omega +2\right)
}{({\beta_1}+2)^2 \left(2 \omega ^2-2 \omega +1\right) ({\beta_1}+4 \omega +1)^2}
\times
\nonumber \\ 
% \displaybreak
&
\times
\Bigl({\beta_1} (\mu  (\omega -1)-1)
+2 \mu  \omega -2 \mu -\omega -1
+({\beta_1}+2) (\omega -1) \log(x^0)
\Bigr)
\,
\left(x^0\right)^{{\beta_1}+1} 
\nonumber \\ 
&
\mbox{}
+
\frac{4 {b_2} (\omega -1) \omega  
\left(
{\beta_2} \left(4 \omega ^2-4 \omega +3\right)+8 \omega ^3+4 \omega +2
\right) 
}{({\beta_2}+2)^2 \left(2 \omega ^2-2 \omega +1\right) ({\beta_2}+4 \omega +1)^2}
\times
\nonumber \\ &
\times
\Bigl({\beta_2} (\mu  (\omega -1)-1)
+2 \mu  \omega -2 \mu -\omega -1
+({\beta_2}+2) (\omega -1) \log(x^0)
\Bigr)
\,
\left(x^0\right)^{{\beta_2}+1} 
\nonumber \\ &
\mbox{}
-
\frac{
{b_1} ({\beta_1}+1) (\omega -1) \left({\beta_1} \left(4 \omega ^2-4 \omega +3\right)+8 \omega ^3+4 \omega +2\right) 
}{({\beta_1}+2) \left(2 \omega ^2-2 \omega +1\right) ({\beta_1}+4 \omega +1)^2}
\,
\left(x^0\right)^{{\beta_1}+1}
\nonumber \\ &
\mbox{}
-
\frac{{b_2} ({\beta_2}+1) (\omega -1) 
\left(
{\beta_2} \left(4 \omega ^2-4 \omega +3\right)+8 \omega ^3+4 \omega +2
\right) 
}{({\beta_2}+2) \left(2 \omega ^2-2 \omega +1\right) ({\beta_2}+4 \omega +1)^2}
\,
\left(x^0\right)^{{\beta_2}+1}
\,
\biggr]\,
\biggr)\, ,
%\end{split}
%\end{equation}
\end{align}
\begin{equation}
\begin{split}
g_{22} = &
-\frac{
\epsilon  
\left(x^0\right)^{4 \omega } 
}{16 (\omega -1)^2 \omega ^2}
\,
\biggl(
4\omega  (\omega -1)  A_{33}
\\ &
\mbox{}
-
\Bigl(
1
-
4 \mu ^2 (\omega -1) \omega -8 \mu  (\omega -1) \omega  \log(x^0)-4 (\omega -1) \omega  \log ^2({x^0})
\Bigr)
\times
\\ &
\times
\Bigl[
A_{22} 
\Bigl(
4 \mu ^2 (\omega -1) \omega-1 +8 \mu  (\omega -1) \omega  \log(x^0)+4 (\omega -1) \omega  \log ^2({x^0})
\Bigr)
\\ &
\mbox{}
-4\omega  (\omega -1)  
\Bigl(
2 A_{23} \bigl(\mu +\log(x^0)\bigr)-A_{33}
\Bigr)
\Bigr]
\biggr)
\\ &
\mbox{}
+
\frac{
1
-
4 \mu ^2\omega  (\omega -1) 
-8 \mu\omega   (\omega -1)  \log(x^0)
-4 \omega  (\omega -1) \log ^2({x^0})
}{4\omega (\omega-1)  }
\,
\left(x^0\right)^{2 \omega },
%\, ,
\end{split}
\end{equation}
%
%\[
\begin{equation}
\begin{split}
g_{23} = &
%\mbox{}
%+
\,
\left(x^0\right)^{2 \omega} \bigl(\mu +\log(x^0)\bigr)
+\mbox{}
\\ &
\mbox{}
+
\frac{
\epsilon  \left(x^0\right)^{4 \omega } 
}{4 (\omega -1) \omega }
\biggl[
\bigl(\mu +\log(x^0)\bigr) 
\Bigl(
4 \mu ^2\omega (\omega -1)  -1 
+8 \mu \omega  (\omega -1)  \log(x^0)
\\ &
\mbox{}
+4\omega (\omega -1)   \log ^2({x^0})
\Bigr)
\,
A_{22}
+
4\omega  (\omega -1)  \bigl(\mu +\log(x^0)\bigr)
\,
A_{33} 
\\ &
\mbox{}
-
\Bigl(
8 \mu ^2 (\omega -1) \omega -1 
+16 \mu  (\omega -1) \omega  \log(x^0)
+8 (\omega -1) \omega  \log ^2({x^0})
\Bigr)
\,
A_{23} 
\biggr]
%\\ &
%\mbox{}
%+
%\left(x^0\right)^{2 \omega} \bigl(\mu +\log(x^0)\bigr)
\, ,
%,\end{equation}
\end{split}
\end{equation}
\begin{equation}
g_{33} = 
-\left(x^0\right)^{2 \omega }
-
\epsilon  
\left(x^0\right)^{4 \omega } 
\biggl(
A_{22} \left(\mu +\log(x^0)\right)^2-2 A_{23} \bigl(\mu +\log(x^0)\bigr)+A_{33}
\biggr)
.\end{equation}
Here \(\omega\) and \(\mu\) are the parameters of the exact solution for the background strong gravitational wave, \(\epsilon\) is the dimensionless smallness parameter (\(\epsilon\ll 1\)), the parameters \(\beta_1\) and \(\beta_2\) are defined by the relations (\ref{CommonBeta1})-(\ref{CommonBeta2}), the parameters \(a_1\), \(a_2\), \(a_3\), \(a_4\), \(b_1\) and \(b_2\) are the constants of integration of the field equations 
and these constants are determined by the initial or boundary conditions. 
The synchronous time function \(\tau\) is defined by the relation
(\ref{CommonTau}). The three functions of the wave variable \(A_{22}(x^0)\), \(A_{23}(x^0)\), and \(A_{22}(x^0)\) share a single equation (\ref{CommonEqForA22A23A33}), which leaves two of these functions arbitrary in the resulting solution.

The determinant of the perturbative gravitational wave metric 
in first order in the 
%dimensionless 
smallness parameter \(\epsilon \) takes the form:
%\[
\begin{align}
\det g_{\alpha\beta} = &
-
\frac{\left(x^0\right)^{4 \omega }}{4\omega (1-\omega)  }
+
\frac{
\epsilon  
\left(x^0\right)^{6 \omega } 
}{16 \omega ^2 (\omega -1)^2}
\biggl[
%-
4 \omega(1-\omega)   \Bigl(2 A_{23} \bigl(\mu +\log(x^0)\bigr)
-
A_{33}
\Bigr)
\nonumber \\
%\]
%\begin{equation}
&
\mbox{}
+
A_{22} \,
\Bigl(
4 \mu ^2 (\omega -1) \omega -1 +8 \mu  (\omega -1) \omega  \log(x^0)+4 (\omega -1) \omega  \log ^2({x^0})
\Bigr)
\biggr].
\end{align}
%.\end{equation}

The solution obtained for the perturbative  gravitational wave metric in the Bianchi IV universe 
%for the most general values of the wave parameters 
allows for seven independent components of the Riemann curvature tensor, whereas the background exact  gravitational wave had  only three independent components of the curvature tensor.

Thus,  perturbative secondary gravitational waves create an additional gravitational wave background that violates isotropy in exact strong gravitational wave models in Bianchi universes and can generate complex tidal accelerations.
The effects under consideration provide an additional mechanism for the formation of local inhomogeneities in the early universe and a mechanism for local spatial isotropization.

\section{Discussion}
\label{sec5}
%We present and successfully apply the proper-time method to construct dynamic perturbative gravitational wave models. 

The paper derives one of the first perturbative analytical models of gravitational waves against the backdrop of an exact solution for a strong gravitational wave.
The stability of the resulting perturbative solutions is demonstrated, which also proves the stability of the basic exact solution for the gravitational wave in the Bianchi IV universe.
Developing such a model for the Bianchi IV universe allows us to further explore the role of perturbative  gravitational waves in the early universe, including the influence of strong and perturbative gravitational waves on the accelerated formation of initial inhomogeneities in dark matter, primordial plasma, and matter. This opens up new possibilities for studying the influence of gravitational waves on the formation of the cosmic microwave background (CMB), taking into account its observed anisotropy, and on the formation of the stochastic gravitational wave background (SGB). This also allows us to estimate the influence of perturbative  gravitational waves on the isotropization of the universe. The constructed analytical model of perturbative   gravitational waves also enables the verification and debugging of numerical methods and computer programs for describing complex gravitational wave models and their influence on astrophysical processes.

\section{Conclusion}
\label{sec6}

The proper-time method for constructing perturbative dynamical gravitational fields is presented.
An analytical perturbative model of gravitational waves based on the exact wave solution of Einstein's equations for the Bianchi IV universe  is constructed. A privileged wave coordinate system and a synchronous time function based on the clock of an observer freely moving (falling) in a background strong gravitational wave were used to construct the models. Compatibility conditions for the vacuum linearized Einstein equations for all admissible parametre values were  obtained and resolved, and the field equations were reduced to systems of ordinary differential equations, whose solutions were found. 
The stability of the resulting perturbative solutions is proven.
This also explicitly demonstrates the stability of the basic exact solution for gravitational waves in the Bianchi IV universe under small perturbations.
The~resulting gravitational wave models have seven independent components of the Riemann curvature tensor, compared to three components for the background exact gravitational wave.

%%%%%%%%%%%%%%%%%%%%%%%%%%%%%%%%%%%%%%%%%%%%%%%%%%%%%

% В конце документа, перед \end{document}:
%\bibliographystyle{unsrtnat} % Сортировка по мере появления, поддержка URL и eprint

%\bibliography{OsetrinListOfCitedPublications-22-02-2022,OsetrinListOfCitedPublications-03-10-2025}

%\bibliography{references}    % Имя вашего .bib файла

\end{document}